\documentclass[preprint,aps,prc,superscriptaddress,showpacs,floatfix]{revtex4}
\usepackage{graphicx}
\begin{document}
\title{Photoproduction of pentaquark cascades from nucleons}
\bigskip
\author{W. Liu}
\affiliation{Cyclotron Institute and Physics Department, Texas A$\&$M
University, College Station, Texas 77843-3366, USA}
\author{C. M. Ko}
\affiliation{Cyclotron Institute and Physics Department, Texas A$\&$M
University, College Station, Texas 77843-3366, USA}

\date{\today}

\begin{abstract}
The cross sections for production of pentaquark $\Xi^+_5$ from the
reaction $\gamma p\to K^0K^0\Xi^+_5$ and $\Xi^{--}_5$ from the reaction
$\gamma n\to K^+K^+\Xi^{--}_5$ are evaluated in a hadronic model
that includes their couplings to both $\Sigma\bar K$ and $\Sigma\bar K^*$.
With these coupling constants determined from the empirical $\pi NN(1710)$
and $\rho NN(1710)$ coupling constants by assuming that
$\Xi^+_5$, $\Xi^{--}_5$, and $N(1710)$ belong to the same antidecuplet
of spin 1/2 and positive parity, and using form factors at strong
interaction vertices similar to those for pentaquark $\Theta^+$ production
in photonucleon reactions, we obtain a cross section of about
0.03-0.6 nb for the reaction $\gamma p\to K^0K^0\Xi^+_5$ and about
0.1-0.6 nb for the reaction $\gamma n\to K^+K^+\Xi^{--}_5$ at photon
 energy $E_\gamma=4.5$ GeV, depending
on the value of the coupling constant $g_{K^*\Sigma\Xi_5}$.
\end{abstract}


\maketitle

\section{Introduction}

The observation of the pentaquark $\Theta^+$ ($uudd\bar s$)
baryon \cite{diakonov} in nuclear reactions induced by photons
\cite{nakano,stepanyan,kubarovsky,barth} and kaons \cite{barmin} has
prompted extensive theoretical studies on both its properties
\cite{prasz,polyakov,walliser,jennings,borisyuk,itzhaki,riska,lipkin,jaffe,hosaka,glozman,zhu,matheus,sugiyama,sasaki,ciskor}
as well as production \cite{liuko,liu,liu4,oh,nam,zhao,hyoto,tsushima}
and decay mechanisms \cite{carl,ma}. Although most models
predict that $\Theta^+$ has spin 1/2 and isospin 0, their predictions
on its parity vary widely.  While the soliton model gives a positive
parity and the lattice QCD study favors a negative parity, the
quark model can give either positive or negative parities, depending
on whether quarks are correlated or not. The parity of $\Theta^+$
affects both the magnitude of its production cross section in these
reactions and the photon asymmetry in photonucleon reactions.
With a positive $\Theta^+$ parity, the production cross section is
almost an order-of-magnitude larger than that for a negative parity
$\Theta^+$ as a result of a smaller $KN\Theta$ coupling constant
in the latter case \cite{liu4,oh,nam}. The positive parity also
leads to a large positive photon asymmetry, which becomes negative if
the $\Theta^+$ parity is negative \cite{zhao,tsushima}.

There are other pentaquark baryons $\Xi^+_5$ ($uuss\bar d$) and
$\Xi^{--}_5$ ($ddss\bar u$) in the same antidecuplet as
$\Theta^+$. Although the $\Xi_5^{--}$ has already been observed
recently in p+p collisions at center-of-mass energy
$\sqrt{s}=17.2$ GeV by the NA49 Collaboration \cite{na49}, its
production in photonucleon reactions has not been studied. In this
paper, we shall use the same hadronic model that was introduced
for studying $\Theta^+$ production in photonucleon reactions
\cite{liuko,liu,liu4} to study the production of $\Xi^+_5$ and
$\Xi^{--}_5$ in these reactions.

This paper is organized as follows. In Section \ref{xi}, we
consider the reactions $\gamma p\to K^0K^0\Xi^+_5$ and $\gamma
n\to K^+K^+\Xi^{--}_5$ for pentaquark $\Xi_5^+$ and $\Xi_5^{--}$
production, respectively, introduce the interaction Lagrangians needed for
evaluate their cross sections, and discuss the coupling constants
and form factors at strong interaction vertices. Results for the
cross sections for these reactions are then shown and discussed in
Section \ref{results}. In Section \ref{summary}, a brief summary
is given. Finally, the formula for evaluating the cross section for
reactions involving two particles in the initial state and three
particles in the final state is derived.

\section{Pentaquark cascade production in photon-nucleon reactions}
\label{xi}

\begin{figure}[ht]
\includegraphics[width=4in,height=3in,angle=0]{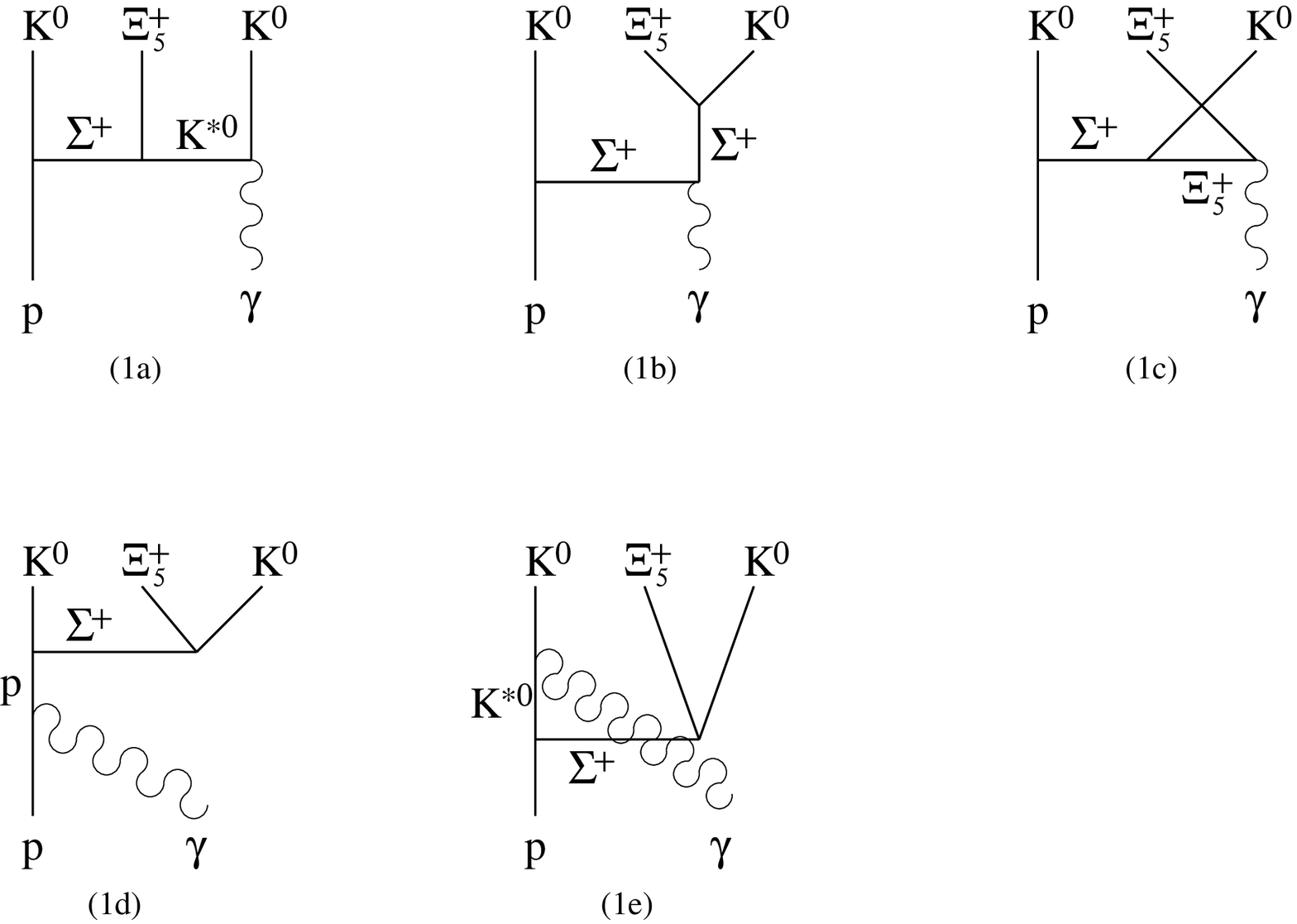}
\caption{Diagrams for pentaquark  $\Xi^+_5$ production from the reaction
$\gamma p\to K^0K^0\Xi^+_5$.}\label{diagram1}
\end{figure}

\begin{figure}[ht]
\includegraphics[width=5in,height=3in,angle=0]{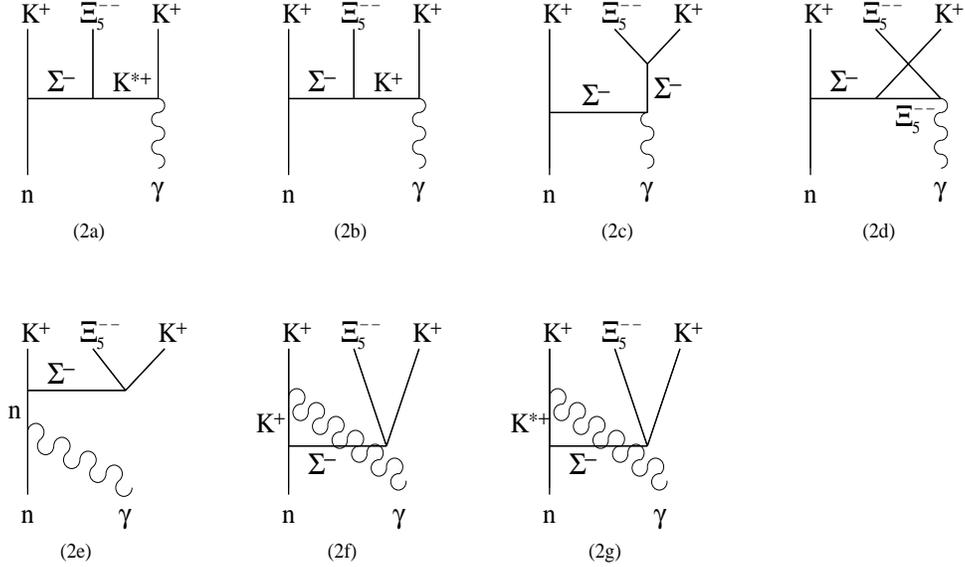}
\caption{Diagrams for pentaquark $\Xi^{--}_5$ production from the reaction
$\gamma n\to K^+K^+\Xi^{--}_5$.} \label{diagram2}
\end{figure}

Possible reactions for pentaquark $\Xi_5$ production in photon-nulceon 
reactions near threshold are $\gamma p\to K^0K^0\Xi^+_5$ and 
$\gamma n\to K^+K^+\Xi^{--}_5$. To evaluate their cross sections, 
we use the hadronic model introduced in Refs.\cite{liuko,liu,liu4} 
for studying $\Theta^+$ production in photon-nucleon reactions. This 
model is a generalization of the SU(3) flavor-invariant Lagrangian for 
the interactions between octet pseudoscalar mesons and baryons 
\cite{li} to include the interactions of $\Theta^+$ with nucleons 
and $K$ as well as $K^*$. Photon in this model is introduced as 
$\rm U_{em}$(1) gauge particle, and the symmetry breaking effects 
are taken into account phenomenologically by using empirical 
hadron masses and coupling constants. In the present study, we 
extend the model to include interactions among the pentaquark 
$\Xi_5$, $\Sigma$, and $K$ or $K^*$. Since only three-point 
interactions appear in the interaction Lagrangians, relevant Feynman 
diagrams that contribute to these reactions are those
shown in Figs. \ref{diagram1} and \ref{diagram2}.

\subsection{Interaction Lagrangians}

The interaction Lagrangians relevant for evaluating the amplitudes shown
in Figs. \ref{diagram1} and \ref{diagram2} are:
\begin{eqnarray}
{\cal L}_{KN\Sigma}&=&ig_{KN\Sigma}\bar N\gamma_5\vec\Sigma\cdot
\vec\tau K + {\rm H.c.},\nonumber\\
{\cal
L}_{K\Sigma\Xi_5}&=&ig_{K\Sigma\Xi}\bar\Xi_5\gamma_5\vec\Sigma
\cdot\vec TK+{\rm H.c.},\nonumber\\
{\cal L}_{K^* N\Sigma}&=&g_{K^* N\Sigma}\bar N\gamma_\mu\vec\Sigma
\cdot\vec\tau K^{*\mu}+{\rm H.c.},\nonumber\\
{\cal
L}_{K^*\Sigma\Xi_5}&=&g_{K^*\Sigma\Xi}\bar\Xi_5\gamma_\mu\vec\Sigma
\cdot\vec TK^{*\mu}+{\rm H.c.},\nonumber\\
{\cal L}_{\gamma NN}&=&-e\bar N\left\{\gamma_\mu\frac{1+\tau_3}{2}A^\mu
-\frac{1}{4m_N}\left[\kappa_p+\kappa_n
+\tau_3(\kappa_p-\kappa_n)\right]\sigma_{\mu\nu}\partial^\nu A^\mu\right\}N
,\nonumber\\
{\cal L}_{\gamma\Sigma\Sigma}&=&-e\bar\Sigma\gamma_\mu Q_1\Sigma A^\mu,
\nonumber\\
{\cal L}_{\gamma KK}&=&ie[\bar KQ_2\partial K-(\partial_\mu\bar K)Q_2K]
A^\mu,\nonumber\\
{\cal L}_{\gamma KK^*}&=&g_{\gamma KK^*}\epsilon_{\alpha\beta\mu\nu}
\partial^\alpha A^\beta[(\partial^\mu\bar K^{*\nu})K+\bar K\partial^\mu
K^{*\nu}].
\end{eqnarray}
In the above, $N$, $\Sigma$, and $\Xi_5$ denote, respectively,
the nucleon isospin doublet, the sigma isospin triplet, and the
pentaquark $\Xi_5$ isospin quartet; $K$ and $K^*$ are the
pseudoscalar and vector kaon isospin doublets, respectively; and
$A_\mu$ denotes the photon. The Pauli matrices are given by
$\vec\tau$, while $\vec T$ is an isopsin transition operator
represented by a $4\times 2$ matrix with matrix elements given by
$\langle\frac{3}{2}m|T_\lambda|\frac{1}{2}
n\rangle=(\frac{3}{2}m|1\lambda\frac{1}{2}n)$, where $\lambda=0,\pm 1$. 
The operators $Q_1={\rm diag}(1,0,-1)$ and $Q_2={\rm diag}(1,0)$ are 
diagonal charge operators, and $\epsilon_{\alpha\beta\mu\nu}$ denotes the
antisymmetric tensor with the usual convention
$\epsilon_{0123}=1$.

For the coupling constants $g_{KN\Sigma}$ and $g_{K^*N\Sigma}$,
their values can be related to $g_{\pi NN}$ and $g_{\rho NN}$,
which has empirical values of $g_{\pi NN}=13.5$ \cite{holzenkamp}
and $g_{\rho NN}=3.25$ \cite{gjanssen}, by SU(3) symmetry, i.e.,
$g_{KN\Sigma}=(1-2\alpha)g_{\pi NN}= -3.78$ and
$g_{K^*N\Sigma}=g_{\rho NN}=3.25$ if we use the ratio
$\alpha=D/(D+F)=0.64$ \cite{adelseck} for the $D-$ and $F-$type
interaction Lagrangians between pseudoscalar mesons and baryons.
Similarly, we have $g_{K\Sigma\Xi_5}=g_{KN\Theta}/\sqrt{2}=2.16$
and $g_{K^*\Sigma\Xi_5}=g_{K^*N\Theta}/\sqrt{2}=1.27$ \cite{kim},
where the values for $g_{KN\Theta}=3.06$ and $g_{K^*N\Theta}=1.8$ have been
determined in \cite{liu4} from the empirical $\pi NN(1710)$ and
$\rho NN(1710)$ coupling constants using the SU(3) symmetry.
Since the sign of $g_{K^*N\Theta}$ relative to $g_{KN\Theta}$
and $g_{KN\Sigma}$ is not fix by the SU(3) symmetry, we shall consider
both signs for the coupling constant $g_{K^*\Sigma\Xi_5}=\pm 1.27$
as well as the case of $g_{K^*\Sigma\Xi_5}=0$.

For photon coupling to nucleon, we include also its interaction
with the anomalous magnetic moment of nucleons with empirical values
of $\kappa_p=1.79$ and $\kappa_n=-1.91$. Since the anomalous magnetic
moment of $\Theta^+$ is not known, we neglect its coupling to photon.

The coupling constant $g_{\gamma KK^*}$ denotes the photon anomalous
parity interaction with kaons and has the dimension of inverse of
energy. Its value is $g_{\gamma K^0K^{*0}}=0.388$ GeV$^{-1}$
and $g_{\gamma K^\pm K^{*\pm}}=0.254$ GeV$^{-1}$ \cite{liu} using the
decay width $\Gamma_{K^{*0}\to K^0\gamma}=0.117$ MeV of $K^{*0}$
and $\Gamma_{K^{*\pm}\to K^\pm\gamma}=0.05$ MeV of $K^{*\pm}$
to kaon and photon \cite{particle}. Although the sign of $g_{\gamma KK^*}$
relative to other coupling constants in the interaction Lagrangians is
not known either, it is not relevant for our study as both constructive
and destructive interferences among the diagrams in
Fig. \ref{diagram1} or Fig.\ref{diagram2} are automatically taken into
account by using different signs for the coupling constant
$g_{K^*\Sigma\Xi_5}$.

\subsection{Amplitudes for the reactions $\gamma p\to K^0K^0\Xi_5^+$
and $\gamma n\to K^+K^+\Xi_5^{--}$}

With the above interaction Lagrangians, we can write the amplitudes
for the five diagrams in Fig.\ref{diagram1} as follows:
\begin{eqnarray}
{\cal M}_{1a}&=&i\sqrt{2}g_{\gamma K^0K^{*0}}
g_{K^*\Sigma\Xi_5}g_{KN\Sigma}\frac{1}{((p_2-p_5)^2-m^2_{K^*})(t-m^2_\Sigma)}
\nonumber\\
&&\times\epsilon_{\alpha\beta\mu\nu}p^\alpha_2 p_5^\beta
\bar\Xi_5(p_4)\gamma^\nu({p\mkern-10mu/}_1-{p\mkern-10mu/}_3+m_\Sigma)
\gamma_5n(p_1)\epsilon^\mu,\nonumber\\
{\cal M}_{1b} & = &-\sqrt{2}eg_{KN\Sigma}g_{K\Sigma\Xi_5}
\frac{1}{(s_1-m^{2}_\Sigma)(t-m^2_\Sigma)}\nonumber\\
&&\times\bar\Xi_5(p_4)({p\mkern-10mu/}_4+{p\mkern-10mu/}_5
-m_\Sigma)\gamma^\mu({p\mkern-10mu/}_1-{p\mkern-10mu/}_3-m_\Sigma)
n(p_1)\epsilon_\mu,\nonumber\\
{\cal M}_{1c}&=&-\sqrt{2}eg_{K\Sigma\Xi_5}g_{KN\Sigma}
\frac{1}{(t-m^2_\Sigma)((p_2-p_4)^2-m^2_{\Xi_5})}\nonumber\\
&&\times\bar\Xi_5(p_4)\gamma^\mu({p\mkern-10mu/}_4-{p\mkern-10mu/}_2
+m_{\Xi_5)}({p\mkern-10mu/}_1-{p\mkern-10mu/}_3-m_\Sigma)
n(p_1)\epsilon_\mu,\nonumber\\
{\cal M}_{1d}&=&\sqrt{2}eg_{K\Sigma\Xi_5}g_{KN\Sigma}
\frac{1}{(s_1-m^2_\Sigma)(s-m^2_N)}\nonumber\\
&&\times\bar\Xi(p_4)({p\mkern-10mu/}_4+{p\mkern-10mu/}_5-m_\Sigma)
({p\mkern-10mu/}_1+{p\mkern-10mu/}_2+m_N)
\left(\gamma^\mu+i\frac{\kappa_p}{2m_N}\sigma^{\mu\nu}p_{2\nu}\right) n(p_1)
\epsilon_\mu,\nonumber\\
{\cal M}_{1e} & = &i\sqrt{2}g_{\gamma K^0K^{*0}}g_{K^*N\Sigma}
g_{K\Sigma\Xi_5}\frac{1}{(s_1-m^{2}_\Sigma)((p_2-p_3)^2-m^2_{K^*})}\nonumber\\
&&\times\bar{\Xi}_5(p_4)\gamma^5({p\mkern-10mu/}_4
+{p\mkern-10mu/}_5+m_\Sigma)\gamma^\nu
n(p_1)\epsilon_{\alpha\beta\mu\nu}
p^\alpha_2p^\beta_3\epsilon^\mu,
\end{eqnarray}
and those for the seven diagrams in Fig.\ref{diagram2} as
\begin{eqnarray}\label{amp2}
{\cal M}_{2a}&=&i\sqrt{2}g_{\gamma K^+K^{*+}}
g_{K^*\Sigma\Xi_5}g_{KN\Sigma}\frac{1}{((p_2-p_5)^2-m^2_{K^*})(t-m^2_\Sigma)}
\nonumber\\
& & \times\epsilon_{\alpha\beta\mu\nu}p^\alpha_2p_5^\beta
\bar\Xi_5(p_4)\gamma^\nu({p\mkern-10mu/}_1-{p\mkern-10mu/}_3+m_\Sigma)
\gamma_5n(p_1)\epsilon^\mu,\nonumber\\
{\cal M}_{2b} &=&-\sqrt{2} eg_{K\Sigma\Xi_5}g_{KN\Sigma}
\frac{1}{((p_2-p_5)^2-m^2_K)(t-m^2_\Sigma)}(2p_5-p_2)^\mu\nonumber\\
&&\times\bar{\Xi}_5(p_4)({p\mkern-10mu/}_1-{p\mkern-10mu/}_3-m_\Sigma)
n(p_1)\epsilon_\mu,\nonumber\\
{\cal M}_{2c} & = &\sqrt{2}eg_{KN\Sigma}g_{K\Sigma\Xi_5}
\frac{1}{(s_1-m^{2}_\Sigma)(t-m^2_\Sigma)}\nonumber\\
&&\times\bar{\Xi}_5(p_4)({p\mkern-10mu/}_4+{p\mkern-10mu/}_5
-m_{\Sigma_5})\gamma^\mu({p\mkern-10mu/}_1-{p\mkern-10mu/}_3-m_\Sigma)
n(p_1)\epsilon_\mu,\nonumber\\
{\cal M}_{2d}&=&2\sqrt{2}eg_{K\Sigma\Xi_5}g_{KN\Sigma}
\frac{1}{(t-m^2_\Sigma)((p_2-p_4)^2-m^2_{\Xi_5})}\nonumber\\
&&\times\bar\Xi_5(p_4)\gamma^\mu({p\mkern-10mu/}_4-{p\mkern-10mu/}_2
+m_{\Xi_5})({p\mkern-10mu/}_1-{p\mkern-10mu/}_3-m_\Sigma)
n(p_1)\epsilon_\mu,\nonumber\\
{\cal M}_{2e}&=&i\sqrt{2}\frac{e\kappa_n}{2m_N}g_{K\Sigma\Xi_5}g_{KN\Sigma}
\frac{1}{(s_1-m^2_\Sigma)(s-m^2_N)}\nonumber\\
&&\times\bar\Xi_5(p_4)({p\mkern-10mu/}_4+{p\mkern-10mu/}_5-m_\Sigma)
({p\mkern-10mu/}_1+{p\mkern-10mu/}_2+m_N)
\sigma^{\mu\nu}p_{2\nu} n(p_1)\epsilon_\mu,\nonumber\\
{\cal M}_{2f} &=&-\sqrt{2} eg_{KN\Sigma}g_{K\Sigma\Xi_5}\frac{1}
{(s_1-m^2_\Sigma)((p_2-p_3)^2-m^2_K)}\nonumber\\
&&\times\bar\Xi_5(p_4)({p\mkern-10mu/}_4+{p\mkern-10mu/}_5-m_\Sigma)
n(p_1)(2p_3-p_2)^\mu\epsilon_\mu,\nonumber\\
{\cal M}_{2g} & = &i\sqrt{2}g_{\gamma K^+K^{*+}}g_{K^*N\Sigma}
g_{K\Sigma\Xi_5}\frac{1}{(s_1-m^{2}_\Sigma)((p_2-p_3)^2-m^2_{K^*})}\nonumber\\
&&\times\bar\Xi_5(p_4)\gamma^5({p\mkern-10mu/}_4+{p\mkern-10mu/}_5
+m_\Sigma)\gamma^\nu n(p_1)\epsilon_{\alpha\beta\mu\nu}
p^\alpha_2p^\beta_3\epsilon^\mu.
\end{eqnarray}
In the above, $p_1$ and $p_2$ denote, respectively, the momenta of nucleon
and photon in the initial states; $p_3$, $p_4$, and $p_5$ are,
respectively, those of the kaon on the left, the cascade, and the kaon
on the right in the final states of the Feynman diagrams in 
Figs. \ref{diagram1} and \ref{diagram2}. We have also introduced 
the following definitions: $s=(p_1+p_2)^2$, $t=(p_1-p_3)^2$, and 
$s_1=(p_4+p_5)^2$.

\subsection{Form factors}

To take into account the effects due to hadron internal structure,
a form factor is introduced to each amplitude, and it is taken to have 
the following form \cite{chung}:
\begin{eqnarray}
F(m_x,m_y)=\left(\frac{\Lambda^4}{\Lambda^4+(q^2_x-m_x^2)^2}\right)
\left(\frac{\Lambda^4}{\Lambda^4+(q^2_y-m_y^2)^2}\right),
\label{form}
\end{eqnarray}
where $q_x$ and $q_y$ are four momenta of the intermediate
off-shell particles with masses $m_x$ and $m_y$ in each diagram,
and $\Lambda$ is the cutoff parameter that characterizes the
off-shell momentum above which hadron internal structure becomes
important. Since the amplitudes for diagram (a) and (e) in Fig.
\ref{diagram1} for the reaction $\gamma p\to K^0K^0\Xi_5^+$ are
individually gauge invariant, including form factors does not
affect their gauge invariance. This is different for the
amplitudes for diagrams (b), (c), and (d) as only their sum is
gauge invariant. Using different form factors for these amplitudes
thus leads to a violation of the gauge invariance. To recover the
gauge invariance for the total amplitude, one can either drop the
gauge violating terms in the amplitude \cite{ohta} or use an
averaged form factor \cite{haberzettl} or other combinations of
form factors \cite{workman} for these terms. For simplicity, we
use here a common form factor, which is given by the average of
the original form factors for diagrams (b), (c), and (d), for all
these three diagrams. A similar consideration is applied to the
diagrams in Fig.\ref{diagram2} for the reaction $\gamma n\to
K^+K^+\Xi_5^{--}$.  In the following, we summarize the form
factors used in the present study:
\begin{eqnarray}\label{contact}
F_{(1a)}&=&F_{(2a)}=F(m_\Sigma,m_{K^*})\nonumber\\
F_{(1b)}&=&F_{(1c)}=F_{(1d)}=\frac{1}{3}[F(m_\Sigma,m_\Sigma)
        +F(m_\Sigma,m_{\Xi_5})+F(m_\Sigma,m_N)]\nonumber\\
F_{(1e)}&=&F_{(2g)}=F(m_\Sigma,m_{K^*})\nonumber\\
F_{(2b)}&=&F_{(2c)}=F_{(2d)}=F_{(2f)}\nonumber\\
        &=&\frac{1}{4}[F(m_\Sigma,m_K)+F(m_\Sigma,m_\Sigma)
        +F(m_\Sigma,m_{\Xi_5})+F(m_\Sigma,m_N)]\nonumber\\
F_{(2e)}&=&F(m_\Sigma,m_N),
\end{eqnarray}
where the subscripts refer to the diagrams in Figs.\ref{diagram1}
and \ref{diagram2}.

The value of the cutoff parameter is taken to be $\Lambda=1.7$ GeV,
which is determined from fitting the measured cross section for
charmed hadron production with three-body final states from
photon-proton reactions at center-of-mass energy of 6 GeV \cite{liu2}
using a similar hadronic model based on SU(4) flavor-invariant
Lagrangians with empirical hadron masses and coupling constants.
This cutoff parameter is smaller than the value of 1.2 GeV used in
our previous study of the reaction $\gamma p\to K^{*0}\Theta^+$
\cite{liu4}, where the form factor is taken to be 
$F=\Lambda^4/(\Lambda^4+m_x^4)$ and the cutoff parameter 
was determined from fitting the cross section for charmed hadron 
production with two-body final states in photon-proton reactions. 

\subsection{Cross sections for $\gamma p\to K^0K^0\Xi_5^+$
and $\gamma n\to K^+K^+\Xi_5^{--}$}

As shown in the Appendix, the total cross section for these reactions
can be expressed as
\begin{eqnarray}
\sigma_{\gamma N\to KK\Xi_5}&=&\frac{1}{(2\pi)^4}\frac{1}{32sp^2_i}
\int\int dt ds_1\nonumber\\
&&\times\frac{k}{2\sqrt{s_1}}\int_0^\pi{\rm sin}\theta_4d
\theta_4\int_0^\pi d\phi_4 |\sum_i{\cal M}_i(\gamma N\to KK\Xi_5)|^2,
\label{cross}
\end{eqnarray}
where $s$, $t$, and $s_1$ were defined after Eq.(\ref{amp2}). We have
also introduced by $p_i$ the momenta of initial-state particles in
their center-of-mass system, and $k$ the momenta of final-state 
particles 4 and 5 in their center-of-mass system. The angles 
$\theta_4$ and $\phi_4$ are polar angles of the three-momentum of 
particle 4 in the center-of-mass system of particles 4 and 5 as 
shown in Fig.\ref{vector} of the Appendix.

\section{results}\label{results}

\begin{figure}[ht]
\includegraphics[width=3.5in,height=4.5in,angle=-90]{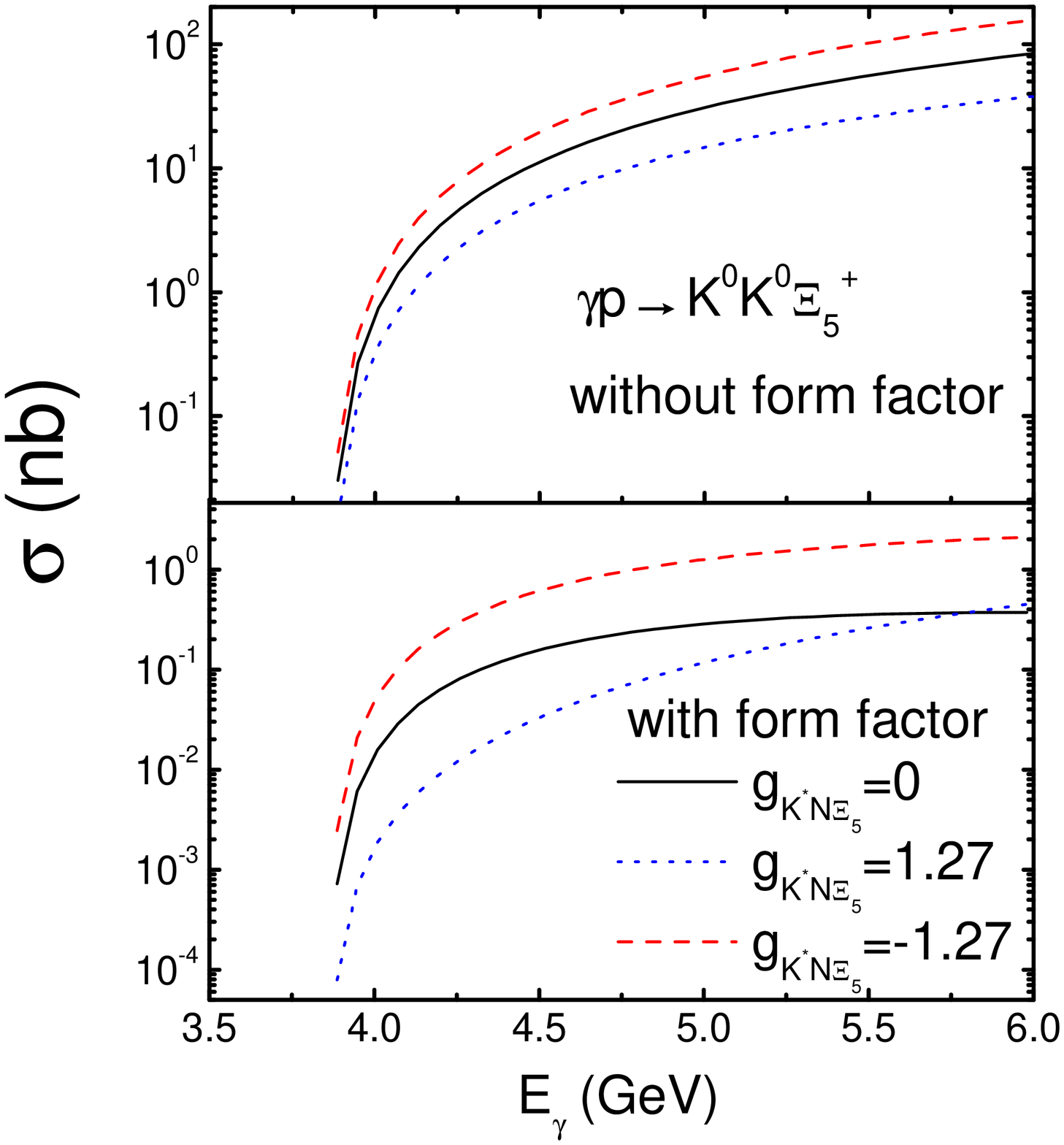}
\caption{(Color online) Total cross section for $\Xi_5^+$ production from 
the reaction $\gamma p\to K^0K^0\Xi_5^+$ as a function of photon energy
and for the coupling constant $g_{K^*\Sigma\Xi_5}$=1.27 (dotted
curve), 0 (solid curve), and -1.27 (dashed curve). Upper and
lower panels are for the cases without and with form factors,
respectively.} \label{cross1}
\end{figure}

\begin{figure}[ht]
\includegraphics[width=3.5in,height=4.5in,angle=-90]{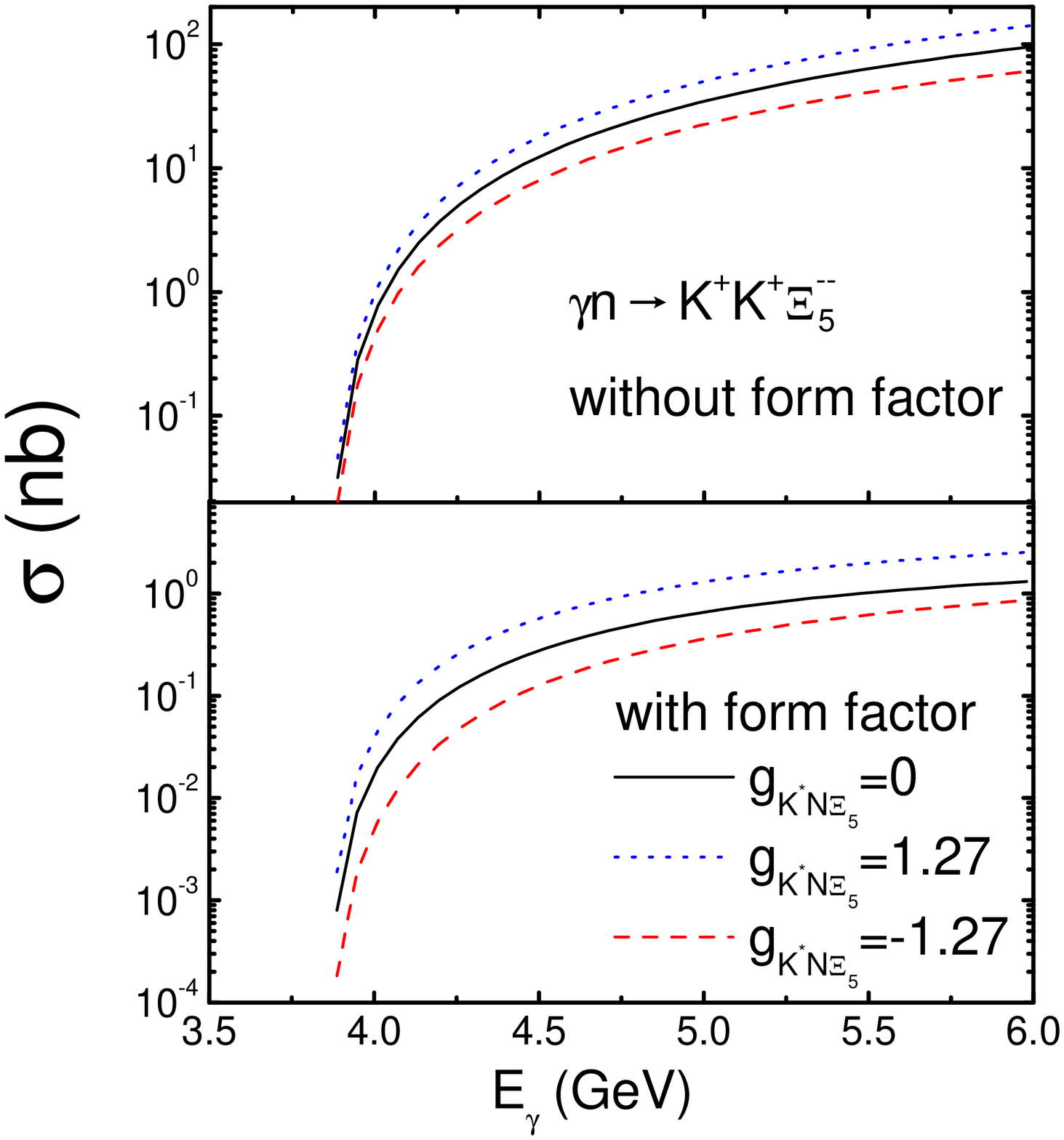}
\caption{(Color online) Total cross section for $\Xi_5^{--}$ production 
from the reaction $\gamma n\to K^+K^+\Xi_5^{--}$ as a function of photon
energy and for the coupling constant $g_{K^*\Sigma\Xi_5}$=1.27
(dotted curve), 0 (solid curve), and -1.27 (dashed curve). Upper and
lower panels are for the case without and with form factors,
respectively.} \label{cross2}
\end{figure}

We first show in the upper panel of Figs.\ref{cross1} and \ref{cross2}
the cross sections for the reactions $\gamma p\to K^0K^0\Xi_5^+$
and $\gamma n\to K^+K^+\Xi_5^{--}$ as functions of photon energy in
the laboratory frame without including form factors at the strong
interaction vertices. Both cross sections are seen to increase
rapidly with increasing photon energy. Since the cross sections
for the case of $g_{K^*\Sigma\Xi_5}$=0 is between those with
$g_{K^*\Sigma\Xi_5}$=1.27 and $g_{K^*\Sigma\Xi_5}$=-1.27, the
contribution from diagrams involving the coupling $g_{K^*\Sigma\Xi_5}$
is less important than those involving the coupling $g_{K\Sigma\Xi_5}$.
In the lower panels, the cross sections with form factors are shown,
and they are much smaller than those without form factors. Depending on the
value of the coupling constant $g_{K^*\Sigma\Xi_5}$, the
cross section for the reaction $\gamma p\to K^0K^0\Xi_5^+$
has values of 0.03-0.6 nb at $E_\gamma=4.5$ GeV, while that for the reaction
$\gamma n\to K^+K^+\Xi_5^{--}$ has values of 0.1-0.6 nb. We note that
these values are significantly smaller that those predicted for
$\Theta^+$ production in photonucleon reactions using the same
hadronic model \cite{liuko,liu,liu4}.

\section{summary}\label{summary}

Using a hadronic model that includes the coupling of pentaquark $\Xi_5$
to usual $\Sigma$ and $K$ or $K^*$, we have evaluated the cross
sections for their production in the reactions $\gamma p\to K^0K^0\Xi_5^+$
and $\gamma n\to K^+K^+\Xi_5^{--}$ by assuming that $\Xi_5$ has spin
1/2 and positive parity. Using coupling constants related to those
for pentaquark $\Theta^+$ couplings to $N$ and $K$ or $K^*$, and also
including form factors at the strong interaction vertices with
empirical cutoff parameters, these cross sections are found
in the range of 0.03-0.6 nb for $\gamma p\to K^0K^0\Xi_5^+$ and
0.1-0.6 nb for $\gamma n\to K^+K^+\Xi_5^{--}$ at photon energy
$E_\gamma=4.5$ GeV. Since the coupling
constant $g_{KN\Theta}$ is much smaller for a negative parity
$\Theta^+$ than for a positive parity one, the coupling constant
$g_{K\Sigma\Xi_5}$ also becomes smaller if $\Xi_5$ has a negative
parity. As a result, the cross sections for producing negative pentaquark
$\Xi_5^+$ and $\Xi_5^{--}$ are much smaller than for positive
ones as in the case of $\Theta^+$ production. Although the cross
sections for photoproduction of pentaquark $\Xi_5$ is an order of magnitude
smaller than those for producing the $\Theta^+$ in these reactions,
measurements of $\Xi_5$ production will be useful for understanding 
the properties of pentaquark baryons.

\section*{Acknowledgments}
This paper was based on work supported in part by the US National
Science Foundation under Grant No. PHY-0098805 and the Welch
Foundation under Grant No. A-1358.

\section*{Appendix}\label{appendix}

In this appendix, we describe the derivation of the cross section
formula shown in Eq.(\ref{cross}) for the reactions
$\gamma p\to K^0K^0\Xi_5^+$ and $\gamma n\to K^+K^+\Xi_5^{--}$
involving two particles in the initial state and three particles
in the final state, i.e., $1+2\to 3+4+5$. It essentially follows the
method given in Ref.\cite{beenakker}.

In the center-of-mass frame of particles 1 and 2, the cross section
for such a reaction generally reads as
\begin{eqnarray}\label{cs}
\sigma_{1+2\to 3+4+5}&=&\frac{1}{(2\pi)^5}\frac{1}{4s^{1/2}p_i}\int\int
d^4p_3d^4p_\Delta \delta(p^2_3-m^2_3)\delta^{(4)}(p_\Delta -p_1+p_3)
\nonumber\\
&&\times \int\int d^4p_4d^4p_5\delta(p^2_4-m^2_4)\delta(p^2_5-m^2_5)
\delta^{(4)}(p-p_4-p_5)\nonumber\\
&&\times |{\cal M} (p_1+p_2\to p_4+p_4+p_5)|^2,
\end{eqnarray}
with $s=(p_1+p_2)^2$ and $p_i$ denoting the momentum of initial
particles in their center-of-mass system.

\begin{figure}[ht]
\includegraphics[width=3.5in,height=3.5in,angle=0]{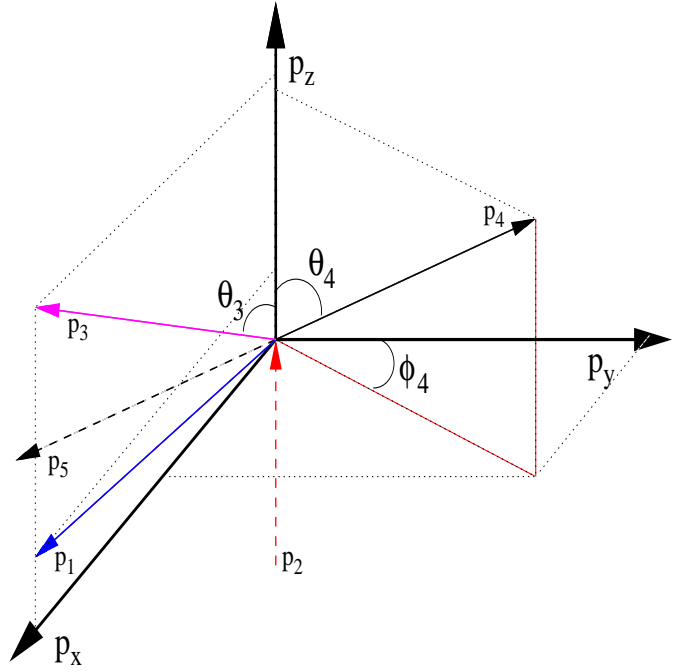}
\caption{
(Color online) Coordinate system used for the three momenta of 
the particles in the reaction $1+2\to 3+4+5$.}
\label{vector}
\end{figure}

To evaluate the phase space integral in Eq.(\ref{cs}), we choose
the coordinate system in the center-of-mass frame of particles 4
and 5 as shown in Fig.\ref{vector}, in which the $z$-axis is along
the three momentum ${\bf p_2}$ of particle 2, ${\bf p_1}$ and
${\bf p_3}$ are in the $x-z$ plane. In terms of the angles shown
in the figure, the four momenta of these particle can be expressed as
\begin{eqnarray}
p_1&=&(E_1,|{\bf p}_3|\sin\theta_3,0,|{\bf p}_3|\cos\theta_3-|
{\bf p}_2|)\nonumber\\
p_2&=&(E_2,0,0,|{\bf p}_2|)\nonumber\\
p_3&=&(E_3,|{\bf p}_3|\sin\theta_3,0, |{\bf p}_3|\cos\theta_3)\nonumber\\
p_4&=&(E_4,|{\bf p}_4|\sin\theta_4\sin\phi_4,
|{\bf p}_4|\sin\theta_4\cos\phi_4,|{\bf p}_4|\cos\theta_4)\nonumber\\
p_5&=&(E_5,-|{\bf p}_4|\sin\theta_4\sin\phi_4,
-|{\bf p}_4|\sin\theta_4\cos\phi_4,-|{\bf p}_4|\cos\phi_4).
\end{eqnarray}

From four-momentum conservation and on-shell constraints, 
the following identities can be derived:
\begin{eqnarray}
E_1&=&\frac{t+s-m^2_2-m^2_3}{2\sqrt{s_1}},
~~E_2=\frac{s_1-t+m^2_2}{2\sqrt{s_1}},
~~E_3=\frac{s-s_1-m^2_3}{2\sqrt{s_1}},\nonumber\\
E_4&=&\frac{s_1+m^2_4-m^2_5}{2\sqrt{s_1}},
~~E_5=\frac{s_1+m^2_5-m^2_4}{2\sqrt{s_1}},
~~\cos\theta_3=\frac{m^2_1-E^2_1+|{\bf p}_3|^2
+|{\bf p}_2|^2}{2|{\bf p}_3||{\bf p}_2|},\nonumber\\
|{\bf p}_2|&=&\sqrt{E^2_2-m^2_2},
~~|{\bf p}_3|=\sqrt{E^2_3-m^2_3},
~~|{\bf p}_4|=\sqrt{E^2_4-m^2_4},
\end{eqnarray}
where $t=(p_1-p_3)^2$ and $s_1=(p_4+p_5)^2$.
The two-body phase-space integral can be rewritten as
\begin{eqnarray}
\int\int d^4p_4&d^4p_5\delta(p^2_4-m^2_4)\delta(p^2_5-m^2_5)
\delta^{(4)}(p-p_4-p_5)=\frac{k}{2\sqrt{s_1}}\int_0^\pi\sin\theta_4
d\theta_4\int_0^\pi d\phi_4,
\end{eqnarray}
where $k$ is the momentum of particle 4 or 5 in their center-of-mass frame.

After integrating over the four-momentum $p_3$, we have
\begin{eqnarray}
\sigma_{1+2\to 3+4+5}&=&\frac{1}{(2\pi)^5}\frac{1}{4s^{1/2}p_i}\int
d^4p_\Delta\delta((p_1-p_\Delta)^2-m^2_3)\nonumber\\
&&\times \frac{k}{2\sqrt{s_1}}\int_0^\pi\sin\theta_4
d\theta_4\int_0^\pi d\phi_4|{\cal M} (p_1+p_2\to p_3+p_4+p_5)|^2.
\end{eqnarray}

In the center-of-mass system of particles 1 and 2, we can rewrite 
$\int d^4p_\Delta$ in terms of $\int ds_1\int dp^2_\Delta=\int ds_1\int dt$
by using the identities
\begin{eqnarray}
s_1&=&m^2_3-m^2_1+m^2_2+2\sqrt{s}E_\Delta,\nonumber\\
p^2_\Delta &=&m^2_3-m^2_1+2E_1E_\Delta-2p_ip_{\Delta z}.
\end{eqnarray}
We then have
\begin{eqnarray}
\sigma_{1+2\to 3+4+5}&=&\frac{1}{(2\pi)^4}\frac{1}{32sp^2_i}\int\int
dt ds_1\nonumber\\
&&\times\frac{k}{2\sqrt{s_1}}\int_0^\pi {\rm sin}\theta_4d
\theta_4\int_0^\pi d\phi_4 |{\cal M} (p_1+p_2\to p_3+p_4+p_5)|^2.
\end{eqnarray}


\begin{thebibliography}{99}

\bibitem{diakonov}D. Diakonov, V. Petrov, and M. Poliakov, Z. Phys.
A {\bf 359}, 305 (1997).
\bibitem{nakano}T. Nakano {\it et al.}, Phys. Rev. Lett. {\bf 91},
012002 (2003).
\bibitem{stepanyan}S. Stepanyan {\it et al.}, CLAS Collaboration,
hep-exp/0307018.
\bibitem{kubarovsky}V. Kubarovsky and S. Stepanyan, hep-ex/0307088.
\bibitem{barth}J. Barth {\it et al.}, SAPHIR Collaboration,
Phys. Lett. B {\bf 572}, 127 (2003).
\bibitem{barmin}V. V. Barmin {\it et al.}, Phys. At. Nucl. {\bf 66},
1715 (2003).
\bibitem{prasz}M. Praszalowicz, Phys. Lett. B {\bf 575}, 234 (2003).
\bibitem{polyakov}M. V. Polyakov and A. Rathke, Eur. Phys. J. A {\bf 18},
691 (2003).
\bibitem{walliser}H. Walliser and V.B. Kopeliovich,
J. Exp. Theor. Phys. {\bf 97}, 433 (2003).
\bibitem{jennings}B. K. Jennings and K. Maltman, hep-ph/0308286.
\bibitem{borisyuk}D. Borisyuk, M. Faber, and A. Kobushkin,
hep-ph/0307370.
\bibitem{itzhaki}N. Itzhaki, I. R. Klebanov, P. Ouyang, and
L. Rastelli, hep-ph/0309305.
\bibitem{riska}Fl. Stancu and D. O. Riska, Phys. Lett. B {\bf 575}, 242
(2003).
\bibitem{lipkin}M. Karliner and H. J. Lipkin, hep-ph/0307243.
\bibitem{jaffe}R. L. Jaffe and F. Wilczek, Phys. Rev. Lett. {\bf 91},
232003 (2003).
\bibitem{hosaka}A. Hosaka, Phys. Lett. B {\bf 55}, 571 (2003).
\bibitem{glozman}L. Y. Glozman, Phys. Lett. B {\bf 575}, 18 (2003).
\bibitem{zhu}S. L. Zhu, Phys. Rev. Lett. {\bf 91}, 232002 (2003).
\bibitem{matheus}R. D. Matheus, F. S. Navarra, M. Nielsen, R. Rodrigues
da Silva, and S. H. Lee, Phys. Lett. B {\bf 578}, 323 (2004).
\bibitem{sugiyama}J. Sugiyama, T. Doi, and M. Oka, hep-ph/0309271.
\bibitem{sasaki}S. Sasaki, hep-lat/0310014.
\bibitem{ciskor}F. Ciskor, Z. Fodor, S. D. Katz, and T. G. Kov\'acs,
JHEP {\bf 0311}, 070 (2003).
\bibitem{liuko}W. Liu and C. M. Ko, Phys. Rev. C {\bf 68}, 045203 (2003);
\bibitem{liu}W. Liu and C.  M. Ko, nucl-th/0309023.
\bibitem{liu4}W. Liu, C. M. Ko, and V. Kubarovsky, nucl-th/0310087,
Phys. Rev. C (to be published).
\bibitem{oh}Y. Oh, H. Kim, and S. H. Lee, hep-ph/0310019, Phys. Rev. D
(to be published); hep-ph/0311054.
\bibitem{nam}S. I. Nam, A. Hosaka, and H. C. Kim, Phys. Lett. B {\bf 579},
43 (2004).
\bibitem{zhao}Q. Zhao, hep-hp/0310350.
\bibitem{hyoto}T. Hyodo, A. Hosaka, and E. Oset, nucl-th/0307105,
Phys. Lett. {\bf B} (to be published).
\bibitem{tsushima}K. Nakayama and K. Tsushima, hep-ph/0311112.
\bibitem{carl}C. E. Garlson, C. D. Carone, H. J. Kwee, and V. Nazaaryan,
Phys. Lett. B {\bf 573}, 101 (2003).
\bibitem{ma}X. Chen, Y. Mao, and B. Q. Ma, hep-ph/0307381.
\bibitem{na49}C. Alt {\it et al.}, NA49 Collaboration, hep-ex/0310014.
\bibitem{li}C. H. Li and C. M. Ko, Nucl. Phys. {\bf A712}, 110 (2002).
\bibitem{holzenkamp}B. Holzenkamp, K. Holinde, and J. Speth, Nucl. Phys.
{\bf A500}, 485 (1989).
\bibitem{gjanssen}G. Janssen, K. Holinde, and J. speth, Phys. Rev. C
{\bf 54}, 2218 (1996).
\bibitem{adelseck}R. A. Adelseck and B. Saghai, Phys. Rev. C {\bf 42},
108 (1990).
\bibitem{kim}Y. Oh, H. Kim, and S. H. Lee, hep-ph/0310117.
\bibitem{particle}Particle Data Group, K. Hagiwara {\it et a.}, Phys.
Rev. D {\bf 66}, 010001 (2002).
\bibitem{chung}W. S. Chung, G. Q. Li, and C. M. Ko, Phys. Lett. B
{\bf 401}, 1 (1997).
\bibitem{ohta}K. Ohta, Phys. Rev. C {\bf 40}, 1335 (1989).
\bibitem{haberzettl}H. Haberzettl, C. Bennhold, T. Mart, and T. Feuster,
Phys. Rev. C {\bf 58}, R40 (1998).
\bibitem{workman}R. M. Davidson and R. Workman, Phys. Rev. C {\bf 63},
058201 (2001).
\bibitem{liu2}W. Liu, S. H. Lee, and C. M. Ko, Nucl. Phys. {\bf A724},
375 (2003).
\bibitem{beenakker}W. Beenakker, H. Kuijf, and W. L. van Neerven,
and J. Smith, Phys. Rev. D {\bf 40}, 54 (1989).

\end{thebibliography}
\end{document}